\documentclass[journal]{IEEEtran}

\usepackage{color}
\usepackage{amsmath, amssymb, bm}
\newcommand{\setsym}[1]{\mathbb{#1}}
\usepackage{cite}

\ifCLASSINFOpdf
\else
\fi
\usepackage{color}

\usepackage{amsmath,amssymb,graphicx,booktabs,url}
\usepackage[utf8]{inputenc}

\begin{document}
%
\title{Recent Progresses in Deep Learning based Acoustic Models (Updated)}


\author{\IEEEauthorblockN{Dong Yu and Jinyu Li} \\
\IEEEauthorblockA{Tencent AI Lab, USA., Microsoft AI and Research, USA.} \\
\IEEEauthorblockA{dongyu@ieee.org, jinyli@microsoft.com}}

%



\IEEEtitleabstractindextext{%
\begin{abstract}
In this paper, we summarize recent progresses made in deep learning based acoustic models and the motivation and insights behind the surveyed techniques. We first discuss acoustic models that can effectively exploit variable-length contextual information, such as recurrent neural networks (RNNs), convolutional neural networks (CNNs), and their various combination with other models. We then describe acoustic models that are optimized end-to-end with emphasis on feature representations learned jointly with rest of the system, the connectionist temporal classification (CTC) criterion, and the attention-based sequence-to-sequence model. We further illustrate robustness issues in speech recognition systems, and discuss acoustic model adaptation, speech enhancement and separation, and robust training strategies. We also cover modeling techniques that lead to more efficient decoding and discuss possible future directions in acoustic model research. \footnote{This is an updated version with latest literature until ICASSP2018 of the paper: Dong Yu and Jinyu Li, ``Recent Progresses in Deep Learning based Acoustic Models,'' vol.4, no.3, IEEE/CAA Journal of Automatica Sinica, 2017.}

\end{abstract}

\begin{IEEEkeywords}
Speech Processing, Deep Learning, LSTM, CNN, Speech Recognition, Speech Separation, Permutation Invariant Training, End-to-End, CTC, Attention Model, Speech Adaptation
\end{IEEEkeywords}}

\maketitle

\IEEEdisplaynontitleabstractindextext

%
\IEEEpeerreviewmaketitle

\section{Introduction}\label{sec:intro}

In the past several years, there has been significant progress in automatic speech recognition (ASR) \cite{PretrainVSFineTune-Yu2010,CD-DNN-HMM-dahl2012,CD-DNN-HMM-SWB-seide2011,DNN4ASR-hinton2012,CNN4ASR-Abdel-Hamid2012,deng2013recent,CNN-Trans-Abdel-Hamid2014,Sak2014long, sak2015learning, CLDNN-sainath2015,DeepCNN-bi2015,TFCNN-mitra2015,TDNN-peddinti2015,VGG-secru2016,Deepspeech2-amodei2015,FSMN-zhang2015,LACE-yu2016,soltau2016neural, HumanParity-Xiong2016,PIT-yu2017,PIT-Kolbak2017}. These progresses have led to ASR systems that surpassed the threshold for adoption in many real-world scenarios and enabled services such as Google Now, Microsoft Cortana, and Amazon Alexa. Many of these achievements are powered by deep learning (DL) techniques. Readers are referred to Yu and Deng 2014 \cite{ASRBook-Yu2014} for a comprehensive summary and detailed description of the technology advancements in ASR made before 2015. 

In this paper, we survey new developments happened in the past two years with an emphasis on acoustic models. We discuss motivations behind, and core ideas of, each interesting work surveyed. More specifically, in Section \ref{sec:am} we illustrate improved DL/HMM hybrid acoustic models that employ deep recurrent neural networks (RNNs) and deep convolutional neural networks (CNNs). These hybrid models can better exploit contextual information than deep neural networks (DNNs) and thus lead to new state-of-the-art recognition accuracy. In Section \ref{sec:end2end} we describe acoustic models that are designed and optimized end-to-end with no or less non-learn-able components. We first discuss the models in which audio waveforms are directly used as the input feature so that the feature representation layer is automatically learned instead of manually designed. We then depict models that are optimized using the connectionist temporal classification (CTC) criterion which allows for a  sequence-to-sequence direct optimization. Following that we analyze models that are built using attention-based sequence-to-sequence model. We devote Section \ref{sec:robust} to discuss techniques that can improve robustness with focuses on adaptation techniques, speech enhancement and separation techniques, and robust training. In Section \ref{sec:efficient} we describe acoustic models that support efficient decoding techniques such as frame-skipping, teacher-student training based model compression, and quantization during training. We propose core problems to work on and potential future directions in solving them in Section \ref{sec:future}. 

\section{Acoustic Models Exploiting Variable-length Contextual Information}\label{sec:am}

The DL/HMM hybrid model \cite{PretrainVSFineTune-Yu2010,CD-DNN-HMM-dahl2012,CD-DNN-HMM-SWB-seide2011,DNN4ASR-hinton2012,CNN4ASR-Abdel-Hamid2012} is the first deep learning architecture that succeeded in ASR and is still the dominant model used in industry. Several years ago, most hybrid systems are DNN based. As reported in \cite{CD-DNN-HMM-SWB-seide2011}, one of the important factors that lead to superior performance in the DNN/HMM hybrid system is its ability to exploit contextual information. In most systems, a window of 9 to 13 frames (left/right context of 4-6 frames) of features are used as the input to the DNN system to exploit the information from neighboring frames to improve the accuracy.

However, the optimal length of  contextual information may vary for different phones and speaking speed. This indicates that using fixed-length context window, as in the DNN/HMM hybrid system, may not be the best choice to exploit contextual information. In recent years people have proposed new models that can exploit variable-length contextual information more effectively. The most important two models use deep RNNs and CNNs.

\subsection{Recurrent Neural Networks} \label{sec:rnn}

Feed-forward DNNs only consider information in a fixed-length sliding window of frames and thus cannot exploit long-range correlations in the speech signal. On the other hand, RNNs can encode sequence history in their internal states, and thus have the potential to predict phonemes based on all the speech features observed up to the current frame. Unfortunately, simple RNNs, depending on the largest eigenvalue of the state-update matrix, may have gradients which either increase or decrease exponentially over time. Hence, the basic RNNs are difficult to train, and in practice can only model short-range effects. 

Long short-term memory (LSTM) RNNs \cite{Hochreiter1997long} were developed to overcome these problems. LSTM-RNNs use input, output and forget gates to control information flow so that  gradients can be propagated in a stable fashion over relatively longer span of time. These networks have been shown to outperform DNNs on a variety of ASR tasks \cite{Graves2013speech, Sak2014long, li2015constructing, Miao15, Miao16}. Note that there is another popular RNN model, called gated recurrent unit (GRU), which is simpler than LSTM but is also able to model the long short-term correlation. Although GRU has been shown effective in several machine learning tasks \cite{chung2014empirical}, it is not widely used in ASR tasks.

At the time step $t$, the vector formulas of the computation of LSTM units can be described as:
\begin{subequations} \label{eq1}
  \begin{align}
  \textbf{i}_t = \sigma ( \textbf{W}_{ix} \textbf{x}_{t} + \textbf{W}_{ih} \textbf{h}_{t-1} + \textbf{p}_{i} \odot \textbf{c}_{t-1} + \textbf{b}_{i}) \label{eq1a}\\
  \textbf{f}_t = \sigma ( \textbf{W}_{fx} \textbf{x}_{t} + \textbf{W}_{fh} \textbf{h}_{t-1} + \textbf{p}_{f} \odot \textbf{c}_{t-1} + \textbf{b}_{f}) \label{eq1b}\\
  \textbf{c}_t = \textbf{f}_t \odot \textbf{c}_{t-1} + \textbf{i}_t \odot \phi( \textbf{W}_{cx} \textbf{x}_{t} + \textbf{W}_{ch} \textbf{h}_{t-1} + \textbf{b}_{c}) \label{eq1c} \\
  \textbf{o}_t = \sigma ( \textbf{W}_{ox} \textbf{x}_{t} + \textbf{W}_{oh} \textbf{h}_{t-1} + \textbf{p}_{o} \odot \textbf{c}_{t} + \textbf{b}_{o}) \label{eq1d} \\
  \textbf{h}_t = \textbf{o}_{t} \odot \phi(\textbf{c}_t) \label{eq1e}
  \end{align}
\end{subequations}
where $\textbf{x}_{t}$ is the input vector. The vectors  $\textbf{i}_t$, $\textbf{o}_t$, $\textbf{f}_t$ are the activation of the input, output, and forget gates, respectively. The $\textbf{W}_{.x}$ and  $\textbf{W}_{.h}$ terms are the weight matrices for the inputs $\textbf{x}_{t}$ and the recurrent inputs $\textbf{h}_{t-1}$, respectively. The $\textbf{p}_{i}$, $\textbf{p}_{o}$, $\textbf{p}_{f}$ are parameter vectors associated with peephole connections. The functions $\sigma$ and $\phi$ are the logistic sigmoid and hyperbolic tangent nonlinearity, respectively. The operation $\odot$ represents element-wise multiplication of vectors. 

It is popular to stack LSTM layers to get better modeling power \cite{Sak2014long}. However, an LSTM-RNN with too many vanilla LSTM layers is very hard to train and there still exists the gradient vanishing issue if the network goes too deep.  This issue can be solved by using either highway LSTM  or residual LSTM. 

In the highway LSTM \cite{HighwayBLSTM-zhang2016}, memory cells of adjacent layers are connected by gated direct links which provide a path for information to flow between layers more directly without decay. Therefore, it alleviates the gradient vanishing issue and enables the training of much deeper LSTM-RNN networks. 

Residual LSTM \cite{zhao2016multidimensional, kim2017residual} uses shortcut connections between LSTM layers, and hence also provides a way to alleviating the gradient vanishing problem. Different from highway LSTM which uses gates to guide the information flow, residual LSTM is more straightforward with the direct shortcut path, similar to Residual CNN \cite{RESNET-he2015} which recently achieves great success in the image classification task.  

Typically, log Mel-filter-bank features are often used as the input to the neural-network-based acoustic model \cite{Mohamed2012understanding, Li12mixedband}. Switching two filter-bank bins will not affect the performance of the DNN or LSTM. However, this is not the case when a human reads a spectrogram: a human relies on both patterns that evolve over time and frequency to predict phonemes.  This inspired the proposal of a 2-D, time-frequency (TF) LSTM \cite{Li15FLSTM, Li16TFLSTM, sainath2016modeling} which jointly scans the speech input over the time and frequency axes to model spectro-temporal warping, and then uses the output activations as the input to the traditional time LSTM. The joint time-frequency modeling provides better normalized features for the upper layer time LSTMs. This has been verified effective and robust to distortion at both Microsoft and Google on large-scale tasks (e.g., Google home \cite{li2017acoustic}). Note that the 2D-LSTM processes along time and frequency axis sequentially, hence it increases computational complexity. In \cite{li2017reducing}, several solutions have been proposed to reduce the computational cost.

Highway LSTM has gates on both the temporal and spatial directions while TF LSTM has gates on both the temporal and spectral directions. It is desirable to have a general LSTM structure that works along all directions. Grid LSTM \cite{kalchbrenner2015grid} is such a general LSTM which arranges the LSTM memory cells into a multidimensional grid. It can be considered as a unified way of using LSTM for temporal, spectral, and spatial computation. Grid LSTM has been studied for temporal and spatial computation in  \cite{hsu2016prioritized} and temporal and spectral computation in \cite{sainath2016modeling}.

Although bi-directional LSTMs (BLSTMs) perform better than uni-directional LSTMs by using the past and future context information \cite{graves2005framewise, Sak2014long}, they are not suitable for real-time systems since the recognition can happen only after the whole utterance has been observed. For this reason, models, such as latency-controlled BLSTM (LC-BLSTM) \cite{HighwayBLSTM-zhang2016} and row-convolution BLSTM (RC-BLSTM), that bridge between uni-directional LSTMs and BLSTMs have been proposed. In these models, the forward LSTM is still kept as is. However, the backward LSTM is replaced by either a backward LSTM with at most $N$-frames of lookahead as in the LC-BLSTM case, or a row-convolution operation that integrates information in the $N$-frames of lookahead. By carefully choosing $N$ we can balance between recognition accuracy and latency. Recently, LC-BLSTM was improved by \cite{Xue2017} to speed up the evaluation and to enable real-time online speech recognition by using better network topology to initialize the BLSTM memory cell states. 

\subsection{Convolutional Neural Networks} \label{sec:cnn}

Another model that can effectively exploit variable-length contextual information is the convolutional neural network (CNN) \cite{CNN-LeCun1995}, in the center of which is the convolution operation (or layer). The input to the convolution operation is usually a three-dimensional tensor \emph{(row, column, channel)} for speech recognition but can be lower or higher dimensional tensors for other applications. Each channel of the input and output of the convolution operation can be considered as a view of the same data. In most setups, all channels have the same size \emph{(height, width)}. 

The filters in the convolution operation are called kernels, which are four-dimensional tensors \emph{(kernel height, kernel width, input channel, output channel)} in our case. There are in total $C_{x} \times C_{v}$ kernels, where $C_{x}$ is the number of input channels and  $C_{v}$ is the number of output channels. The kernels are applied to local regions called receptive fields in an input image along all channels.  The value after the convolution operation is
\begin{eqnarray}
\upsilon_{ij\ell}\left(\mathbf{K,X}\right) & = & \sum_{n}\mathrm{vec}\left(\mathbf{K}_{n\ell}\right)\cdot\mathrm{vec}\left(\mathbf{X}_{ijn}\right),
\end{eqnarray}
for each output channel $\ell$ and input slice $\left(i,j\right)$ (the $i$-th step along the vertical direction and $j$-th step along the horizontal direction), where $\mathbf{K}_{n\ell}$ of size $(H_k, W_k)$ is a kernel matrix associated with input channel $n$ and output channel $\ell$ and has the same size as the input image patch $\mathbf{X}_{ijn}$ of channel $n$, $\mathrm{vec}(\dot)$ is the vector formed by stacking all the columns of the matrix, and $\cdot$ is the inner product of two vectors. It is obvious that each output pixel is a weighted sum of all pixels across all channels in an input patch. Since each input pixel can be considered as a weak patten detector, each output pixel is just a boosted detector exploiting all information in the input patch. 

The kernel is shared across all input patches and moves along the input image with strides $S_{r}$ and $S_{c}$ at the vertical and horizontal direction, respectively. When the strides are larger than 1, the convolution operation also subsamples, in additional to convolving, the input image and leads to a lower-resolution image that is less sensitive to the small pattern shift inside the input patch. The translational invariance can be further improved when some kind of aggregation operations are applied after the convolution operation. Typical aggregation operations are max-pooling and average-pooling. The aggregation operations often go with subsampling to reduce resolution. Due to the built-in translational invariability CNNs can exploit variable-length contextual information along both frequency and time axes. It is obvious that if only one convolution layer is used the translational variability the system can tolerate is limited. To allow for more powerful exploitation of the variable-length contextual information, convolution operations (or layers) can be stacked.

The time delay neural network (TDNN) \cite{TDNN-lang1990} was the first model that exploits multiple CNN layers for ASR. In this model, convolution operations are applied to both time and frequency axes. However, the early TDNNs are neural network only solution that do not integrate with HMMs and are hard to be used in large vocabulary continuous speech recognition (LVCSR).

After the successful application of DNNs to LVCSR, CNNs were reintroduced under the DL/HMM hybrid model architecture \cite{CNN4ASR-Abdel-Hamid2012,CNN-Attention-Abdel-Hamid2013,DeepCNN-Sainath2013,CNN-Trans-Abdel-Hamid2014,DeepCNN-Sainath2013,DeepCNN-bi2015,VGG-secru2016,LACE-yu2016,CNN-Dilated-sercu2016}. Because HMMs in the hybrid model already have strong ability to handle variable-length utterance problem in ASR, CNNs were reintroduced initially to deal with variability at the frequency axis only \cite{CNN4ASR-Abdel-Hamid2012,CNN-Attention-Abdel-Hamid2013,DeepCNN-Sainath2013,CNN-Trans-Abdel-Hamid2014}. The goal was to improve robustness against vocal tract length variability between different speakers. Only one to two CNN layers were used in these early models, stacked with additional fully-connected DNN layers. These models have shown around 5\% relative recognition error rate reduction compared to the DNN/HMM systems \cite{CNN-Trans-Abdel-Hamid2014}. Later, additional RNN layers, e.g., LSTMs, were integrated into the model to form so called CNN-LSTMM-DNN (CLDNN) \cite{CLDNN-sainath2015} and CNN-DNN-LSTM (CDL) architectures. The RNNs in these models can help to exploit the variable-length contextual information since CNNs in these models only deal with frequency-axis variability. CLDNN and CDL both achieved additional accuracy improvement over CNN-DNN models.

Researchers quickly realized that dealing with variable-length utterance is different from exploiting variable-length contextual information. TDNNs, which convolve along both the frequency and time axes and thus exploit variable-length contextual information, attracted new attentions, this time under the DL/HMM hybrid architecture \cite{CNNtemporal-toth2015,TDNN-peddinti2015} and with variations such as row convolution \cite{Deepspeech2-amodei2015} and feedforward sequential memory network (FSMN) \cite{FSMN-zhang2015}. Similar to the original TDNNs, these models stack several CNN layers along the frequency and time-axis, with a focus on the time-axis, to account for speaking rate variation. But unlike the original TDNNs, the TDNN/HMM hybrid systems can recognize large vocabulary continuous speech very effectively.

More recently, primarily motivated by the successes in image recognition, various architectures of deep CNNs \cite{TFCNN-zhao2015,VGG-secru2016,LACE-yu2016,CNN-Dilated-sercu2016} have been proposed and evaluated for ASR. The premise is that spectrograms can be seen as images with special patterns from which experienced people can tell what has been said. In deep CNNs, each higher layer is a weighted sum of nonlinear transformation of a window of lower layers and thus covers longer contexts and operates on more abstract patterns. Lower CNN layers capture local simple patterns while higher CNN layers detect broader, abstract, and more complicated patterns. Smaller kernels combined with more layers allow deep CNNs to exploit longer-range dependency information along both time and frequency axes more effectively. Empirically deep CNNs are compatible to BLSTMs \cite{HumanParity-Xiong2016}, which in turn outperform unidirectional LSTMs. However, unlike BLSTMs which suffer from long latency, because you need to wait for the whole utterance to finish to start decoding, and cannot be deployed in real-time systems, deep CNNs have limited latency and are better suited for real-time systems if the computation cost can be controlled.

Training and evaluation of deep CNNs is very time consuming, esp. if we treat each window of frames independently, under which condition there are significant duplication of computations. To speedup the computation we can treat the whole utterance as a single input image and thus reuse the intermediate computation results. Even better, if the deep CNN is designed so that the stride at each layer is long enough to cover the whole kernel, similar to the CNNs with layer-wise context expansion and attention (LACE) \cite{LACE-yu2016}. Such model, called dilated CNN \cite{CNN-Dilated-sercu2016}, allows to exploit longer-range information with less number of layers and can significantly reduce the computational cost. Dilated CNN has outperformed other deep CNN models on the switchboard task \cite{CNN-Dilated-sercu2016}. 

Note that deep CNNs can be used together with RNNs and under frameworks such as connectionist temporal classification (CTC) that we will discuss in Section \ref{sec:ctc}.

\section{Acoustic Models with End-to-end Optimization} \label{sec:end2end}

The models discussed in the previous section are DNN/HMM hybrid models in which the two components DNN and HMM are usually optimized separately. However, speech recognition is a sequential recognition problem. It is not surprising that better recognition accuracy may be achieved if all components in a model are jointly optimized. Even better, if the model can remove all manually designed components such as basic feature representation and lexicon design.  

\subsection{Automatically Learned Audio Feature Representation} \label{sec:wave}

It is always arguable that the manually-designed log Mel-filter-bank feature is  optimal for speech recognition. Inspired by the end-to-end processing in the machine learning community, there are always efforts \cite{jaitly2011learning, palaz2014estimating, tuske2014acoustic, sainath2015learning} trying to replace the Mel-filter-bank extraction by directly learning filters using a network to process the raw speech waveforms and training it jointly with the recognizer network. Among these efforts, the CLDNN \cite{CLDNN-sainath2015} on raw waveform work \cite{sainath2015learning} seems to be more promising as it got slight gain over the log Mel-filter-bank feature while the other works didn't. More importantly, it serves a good foundation of the multichannel processing with raw waveforms. 

The most critical thing for raw waveform processing is using a representation that is invariant to small phase shift because the raw waveforms are perceptually identical if the only  difference is a small phase shift. To achieve the phase invariance, a time convolutional layer is applied to the raw waveform and then pooling is done over the entire time length of the time-convolved output signal. This process reduces the temporal variation (hence phase invariant) and is very similar to the Gammatone filterbank extraction. The pooled outputs can be considered as the filter-bank outputs, on which the standard CLDNN \cite{CLDNN-sainath2015} is applied. Interestingly, the same idea was recently applied to anti-spoofing speaker verification task with significant gain \cite{Heinrich17}. 

With the application of deep learning models, now the ASR systems on close-talking scenario perform very well. The research interest is shifted to the far-field ASR which needs to handle both additive noise and reverberation. The current dominant approach is still using the traditional beamforming method to process the waveforms from multiple microphones, and then inputting the beamformed signal into the acoustic model \cite{yoshioka2015ntt}. Efforts have also been made to use deep learning models to perform beamforming and to jointly train the beamforming and recognizer networks \cite{xiao2016deep, sainath2015speaker, sainath2016factored, sainath2017multichannel}. In \cite{xiao2016deep}, the beamforming network and recognizer networks were trained in a sequence by first training the beamforming network, then training the recognizer network with the beamformed signal, and finally jointly training both networks. 

In \cite{sainath2015speaker, sainath2016factored, sainath2017multichannel}, both networks were jointly trained in a more end-to-end fashion by extending the aforementioned CLDNN to raw waveform. In the first layer, multiple time convolution filters are used to map the raw waveforms from multiple microphones into a single time-frequency representation \cite{sainath2015speaker}. Then the output is passed to the upper layer CLDNN for phoneme classification. Later, the joint network is improved by factorizing the spatial and spectral selectivity of bottom layer network into a spatial filtering layer and a spectral filtering layer. The factored network brings accuracy improvement with the cost of increasing computational cost, which later was reduced by converting the time-domain convolution into the frequency-domain product \cite{Variani2016}. Later, the CNN layer was replaced by the 2D-LSTM layer \cite{sainath2016modeling} to improve the robustness, and Google home was built with this end-to-end system \cite{li2017acoustic}.

\subsection{Connectionist Temporal Classification} \label{sec:ctc} 

Speech recognition task is a sequence-to-sequence task, which maps the input waveform to a final word sequence or an intermediate phoneme sequence. What the acoustic modeling cares is the output of word or phoneme sequence, instead of the frame-by-frame labeling which the traditional cross entropy (CE) training criterion cares. Hence, the Connectionist Temporal Classification (CTC) approach \cite{sak2015learning, sak2015fast, Senior15} was introduced to map the speech input frames into an output label sequence. As the number of output labels is smaller than that of input speech frames, CTC path is introduced to force the output to have the same length as the input speech frames by adding blank as an additional label and allowing repetition of labels. 

Denote $\bf{x}$ as the speech input sequence, $\bf{l}$ as the original label sequence, and  $B^{-1}(\bf{l})$ represents all the CTC paths mapped from $\bf{l}$. Then, the CTC loss function is defined as the sum of negative log probabilities of correct labels as
\begin{equation}
L_{CTC} = - ln P( \bf{l}|\bf{x} ),
\end{equation}
where 
\begin{equation} \label{eq:Prob_CTC}
P( {\bf{l}|\bf{x}} ) = \sum_{\bf{z} \in B^{-1}(\bf{l})} P( \bf{z} | \bf{x} )
\end{equation}
With the conditional independent assumption, $P( \bf{z} | \bf{x} )$ can be decomposed into a product of posterior from each frame as
\begin{equation}
P( {\bf{z} | \bf{x}} ) = \prod_{t=1}^T P( z_t | \bf{x}).
\end{equation}

In \cite{sak2015fast}, CTC with context-dependent phone output units have been shown to outperform CTC with monophone \cite{sak2015learning}, and to perform in par with the LSTM model with cross entropy criterion when training data is large enough. One attractive characteristics of CTC is that because the output unit is usually phoneme or unit larger than phoneme, it can take larger input time step instead of single 10ms frame shift. For example, in \cite{sak2015fast}, three 10ms frames are stacked together as the input to CTC models. By doing so, the acoustic score evaluation and decoding happen every 30ms, 3 times faster than the traditional systems that operate on 10ms frame shift.

The most attractive characteristics of CTC is that it provides a path to end-to-end  optimization of acoustic models. In the deep speech \cite{hannun2014deep, Deepspeech2-amodei2015} and EESEN \cite{miao2015eesen, miao2016empirical} work, the end-to-end speech recognition system is explored to directly predict characters instead of phonemes, hence removing the need of using lexicons and decision trees which are the building blocks in \cite{sak2015learning, sak2015fast, Senior15}. This is one step toward removing expert knowledge when building an ASR system. Another advantage of character-based CTC is that it is more robust to the accented speech as the graphoneme sequence of words is less affected by accents than the phoneme pronunciation  \cite{sak2017multi}. Other output units that are larger than characters but smaller than words have also been studied \cite{zweig2017}. 

It is a design challenge to determine the basic output unit to use for CTC prediction. In all the aforementioned works, the decomposition of a target word sequence into a sequence of basic units is fixed. However, the pre-determined fixed decomposition is not necessarily optimal. In \cite{liu2017gram}, gram-CTC was proposed to automatically learn the most suitable decomposition of target sequences. Gram-CTC is based on characters, but allows to output variable number of characters (i.e., gram) at each time step. This not only boosts the modeling flexibility  but also improves the final ASR system accuracy. 

\begin{figure}[h]
\begin{center}
\includegraphics[width=0.5\textwidth]{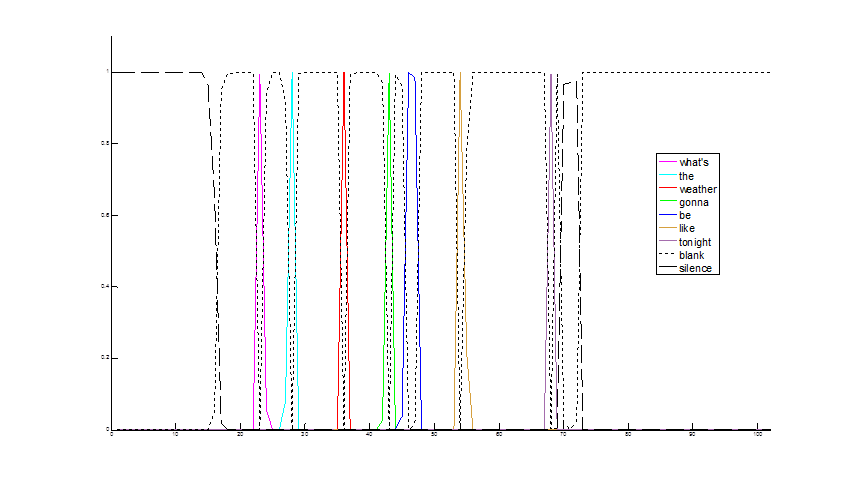}
\end{center}
\caption{An example of word CTC. }
\label{fig:wordCTC}
\end{figure}

As the goal of ASR is to generate a word sequence from the speech waveform, word unit is the most natural output unit for network modeling. In \cite{sak2015fast},  CTC with word output targets was explored but the accuracy is far from the phoneme-based CTC system. In \cite{soltau2016neural}, it was shown that by using 100k words as the output targets and by training the model with 125k hours of data, the CTC system with word units can beat the CTC system with phoneme unit. Figure \ref{fig:wordCTC} gives an example of the posterior output of word CTC. In the figure, the units with the maximum posterior values  are blanks and silences at most of time steps. All other posterior spikes come from word units. Hence, the ASR task becomes very simple: the output word sequence is constructed by taking the words corresponding to posterior spikes. No language model or complex decoding process is involved. 
A big challenge in the word-based CTC is the out-of-vocabulary (OOV) issue. In \cite{sak2015fast, soltau2016neural, audhkhasi2017direct}, only the most frequent words in the training set were used as targets whereas the remaining words were just tagged as OOVs. All these OOV words can neither be further modeled nor be recognized during evaluation.  

To solve this OOV issue in the word-based CTC,  a hybrid CTC was proposed  \cite{Li17CTCnoOOV} to use the output from the word-based CTC as the primary ASR result and consults a character-based CTC at the segment level where the word-based CTC emits an OOV token. In \cite{audhkhasi2018building}, a spell and recognize model was used to learn to first spell a word and then recognize it. Whenever an OOV is detected, the decoder consults the character sequence from the speller. In \cite{Li17CTCnoOOV, audhkhasi2018building}, the displayed hypothesis is more meaningful than OOV to users. However, both methods cannot improve the overall recognition accuracy too much due to the two-stage (OOV-detection and then character-sequence-consulting) process. In \cite{Li18CTCnoOOV}, a better solution was proposed by decomposing the OOV word into a mixed-unit sequence of frequent words and characters at the training stage. During testing, the whole system uses greedy decoding to generate hypotheses in a single step without the need of using the two-stage processing. Combined with attention modeling for CTC \cite{Das18CTCAttention}, such an end-to-end model without using any LM or complex decoder can significantly outperform the traditional context-dependent phoneme CTC which has strong LM and decoder.

Compared to traditional cross-entropy training of LSTM, CTC is harder to train. First, the network initialization is very important. In \cite{sak2015learning}, the LSTM network for CTC training was initialized from the LSTM network trained with cross entropy criterion. This can be circumvented by using very large amount of training data which also helps to prevent overfitting \cite{sak2015fast}.  If the CTC network is randomly initialized, when presented very difficult samples, the CTC network tends to be very hard to train. In \cite{Deepspeech2-amodei2015}, a learning strategy called SortaGrad was proposed by first presenting the CTC network with shorter utterances (easy samples) and then longer utterances (hard samples) in the first training epoch. In the later epochs, the utterances are given to CTC network randomly. This significantly improves the convergence of CTC training. 

The spike patterns in Figure \ref{fig:wordCTC} is general to CTC modeling with any basic units. Therefore, at the time steps that blank symbol dominates, it may be redundant to do the search as no information is provided. Given this observation, phone synchronous decoding \cite{chen2017phone} was proposed by skipping the search of blank-dominated time steps during CTC decoding. 2-3 times speedup was obtained without accuracy loss.  

Note that the occurrence of spikes in CTC usually has a delay compared to the ground-truth location of the symbol. Such a delay introduces latency during the runtime decoding which is not desirable to the systems with realtime requirement. Therefore, a delay-constrained training was proposed in \cite{Senior15} by restricting the search paths used in the forward-backward process during CTC training to those in which the delay between CTC labels and the ground-truth alignment does not exceed a threshold. This constraint degrades the CTC training a little, but the loss was recovered after sequence training. 

The frame-independence assumption in CTC is most criticized. There are several attempts to improve CTC modeling by relaxing or removing such assumption. In \cite{Das18CTCAttention}, attention modeling was directly integrated into the  CTC framework by using time convolution features, non-uniform attention, implicit
language modeling, and component attention. Such an attention CTC model relaxes the frame-independence assumption by working on model hidden layers without changing the CTC objective function and training process, hence enjoying the simplicity of CTC modeling.  On the other hand, RNN transducer \cite{Graves-RNNSeqTransduction}
and RNN aligner \cite{sak2017recurrent} extend CTC modeling by changing the objective function and the training process to remove the frame-independence assumption of CTC. Specifically, RNN transducer was shown very effective \cite{battenberg2017exploring, rao2017exploring} as it incorporates acoustic model with its encoder, language model with its prediction network, and decoding process with its joint network.

Inspired by the CTC work, lattice-free maximum mutual information (LFMMI) \cite{povey2016} was recently proposed to train deep networks from scratch without initializing from cross-entropy networks. This single-step training has great advantage over current popular two-step training: first cross-entropy training and then sequence training. Lots of efforts have been developed to make LFMMI work, including a topology that the first frame of a phoneme has a different label than the remaining frames; a phoneme n-gram language model used to create denominator graph; a time-constraint similar to the delay-constrain used in CTC; several regularization methods to reduce overfitting; stacking multiple input frames as what CTC does; etc. LFMMI has been proven effective on tasks with different scale and underlying models. 

Although there are lots of models proposed in recent years, clearly there is a major AM developing line from DNN to LSTM (temporal modeling) and then to CTC (end-to-end modeling). Although some models can achieve similar performance as CTC when modeling phonemes, they may not fit the trend of end-to-end modeling very well as these models still require expert knowledge to design and need components such as language model and lexicon. 

\subsection{Attention-based Sequence-to-Sequence Models} \label{sec:attention}

Attention-based sequence-to-sequence model is another end-to-end model \cite{bahdanau2016end, chan2016listen}. It roots from the successful attention model in machine learning \cite{bahdanau2014neural, mnih2014recurrent} which extends the encoder-decoder framework \cite{cho2014learning} using an attention decoder.
The attention model calculates the probability at step $i$ as  
\begin{equation} \label{eq:Prob_attention} 
P( {\bf{l}|\bf{x}} ) = \prod_u P(l_u | {\bf{x}} , {\bf{l}}_{1:u-1}),
\end{equation}
with
\begin{align}	
 P(l_u | {\bf{x}} , {\bf{l}}_{1:u-1}) & =  AttentionDecoder({\bf{h}}, {\bf{l}}_{1:u-1}), \label{eq:attendecoder} \\
{\bf{h}} & = Encoder ({\bf{x}}). \label{eq:encoder}
\end{align}
The training criterion is to minimize $- ln P( {\bf{l}}|{\bf{x}} )$. 

\begin{figure}[h]
\begin{center}
\includegraphics[width=0.5\textwidth]{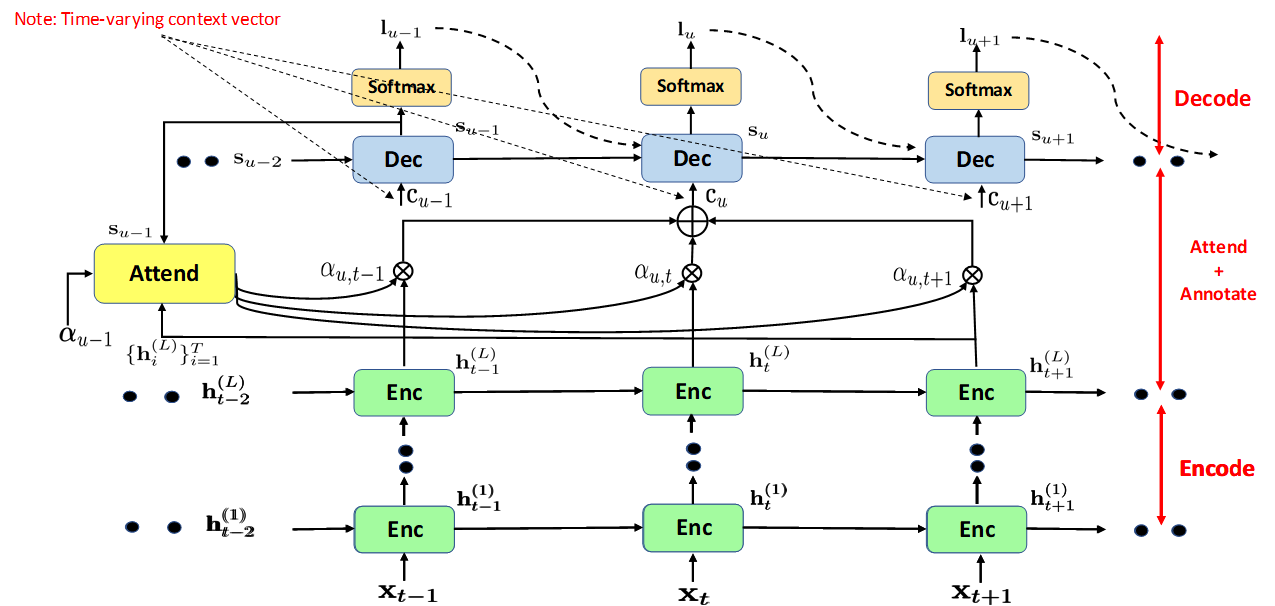}
\end{center}
\caption{The flowchart of attention-based model. }
\label{fig:attention}
\end{figure}

The flowchart of attention-based model is given in Figure \ref{fig:attention}. Different from the encoder in \cite{cho2014learning} which only takes the hidden vector of last time step, the encoder in Eq. \eqref{eq:encoder} transforms the whole speech input sequence $\bf{x}$ to a high-level hidden vector sequence ${\bf{h}} = ( {\bf{h}}_1, {\bf{h}}_2, ......, {\bf{h}}_L), L\leq T$.  Then, at each  step in generating an output label $l_i$, an attention mechanism in Eq. \eqref{eq:attendecoder} selects/weights the hidden vector sequence ${\bf{h}}$ so that the most related hidden vectors are used for the prediction. Comparing Eq. \eqref{eq:Prob_attention} with Eq. \eqref{eq:Prob_CTC}, we can see the attention-based model doesn't have the frame-independence assumption imposed by CTC, which is the advantage of the attention model. 

The AttentionDecoder network has three components: a multinomial distribution generator \eqref{eq:RNNED-generate}, an RNN decoder \eqref{eq:RNNED-recurrent}, and an attention network \eqref{eq:RNNED-annotate}-\eqref{eq:RNNED-locfeat} as follows:
\begin{align}
{\bf{l}}_{u} &= \text{Generate}({\bf{l}}_{u-1}, {\bf{s}}_{u}, {\bf{c}}_{u}), \label{eq:RNNED-generate} \\
{\bf{s}}_{u} &= \text{Recurrent}({\bf{s}}_{u-1}, {\bf{l}}_{u-1}, {\bf{c}}_{u}), \label{eq:RNNED-recurrent} \\
{\bf{c}}_{u} &= \text{Annotate}({\bm\alpha_{u}}, {\bf{h}}) = \sum_{t=1}^{T} \alpha_{u,t} {\bf{h}}_{t} \label{eq:RNNED-annotate} \\
{\bm\alpha}_{u} &= \text{Attend}({\bf{s}}_{u-1}, {\bm\alpha}_{u-1}, \bf{h}). \label{eq:RNNED-attend}
\end{align}

Here, ${\bf{l}}_{u} \in \setsym{U}^{K}$, ${\bf{h}}_{t}, {\bf{c}}_{u} \in \setsym{R}^{n}$, $\bm\alpha_{u} \in \setsym{U}^{T}$, and for simplicity ${\bf{s}}_{u} \in \setsym{R}^{n}$. $\text{Generate}(.)$ is a feedforward network with a softmax operation  generating the probability of the target output $p({\bf{l}}_{u}|{\bf{l}}_{u-1}, {\bf{s}}_{u}, {\bf{c}}_{u})$.  Recurrent(.) is an RNN decoder operating on the output time axis indexed by $u$ and has hidden state ${\bf{s}}_{u}$. Annotate(.) computes the context vector ${\bf{c}}_{u}$ (also called the soft alignment) using attention probability vector $\bm\alpha_{u}$. Attend(.) computes the attention weight $\alpha_{u,t}$ using a single layer feedforward network as,
\begin{align}
e_{u,t} &= \text{Score}({\bf{s}}_{u-1}, \bm\alpha_{u-1}, {\bf{h}}_{t}), \label{eq:RNNED-score} \\
\alpha_{u, t} &= \frac{ \text{exp}(e_{u, t}) } { \sum_{t^{\prime}=1}^{T} \text{exp}(e_{u, t^{\prime}}) }, \label{eq:RNNED-normalizedscore}
\end{align}
where $e_{u, t} \in \setsym{R}$. $\text{Score}(.)$ can either be content or hybrid. It is computed using,
\begin{align}
\hspace{-2mm} e_{u, t} &= \begin{cases}
{\bf{v}}^{T}\text{tanh}({\bf{U}} {\bf{s}}_{u-1} + {\bf{W}} {\bf{h}}_{t}  +  {\bf{b}}), \ \mbox{(content)} \\
{\bf{v}}^{T}\text{tanh}({\bf{U}} {\bf{s}}_{u-1} + {\bf{W}} {\bf{h}}_{t}  + {\bf{v}} {\bf{f}}_{u,t} + {\bf{b}}), \ \mbox{(hybrid)}
\end{cases} \label{eq:RNNED-ContentHybrid} \\
&\text{where,} \quad {\bf{f}}_{u,t} = {\bf{f}} \ast \bm\alpha_{u-1} \label{eq:RNNED-locfeat}.
\end{align}
The operation $\ast$ denotes convolution. ${\bf{U}}, {\bf{W}}, {\bf{v}}, {\bf{f}}, {\bf{b}}$ are trainable attention parameters .

The attention-based model is even harder to train than the CTC model. There are plenty of tricks to be applied. For example, the vanilla attention-based model is highly complex during training if all the hidden vectors at all time steps are used in Eq. \eqref{eq:attendecoder}. Therefore, windowing method is used in \cite{bahdanau2016end} to reduce the number of candidates used in attention decoder. In \cite{chan2016listen}, a pyramid structure is used in the encoder network so that only $L$ high-level hidden vectors are generated instead of $T$ hidden vectors from all the input time steps. Due to the high complexity and slow speed of training, the majority of attention-based works were majorly done at Google \cite{chan2016listen}, compared to the CTC works reported from many sites. However, recently Google significantly advanced the research of attention-based model by 
\begin{itemize}
	\item modeling with word piece units \cite{schuster2012japanese} which are more stable and helpful to LM modeling; 
	\item including scheduled sampling \cite{bengio2015scheduled} which feeds the previous predicted label instead of the ground truth during training so that training and testing are consistent;
	\item having multi-head attention \cite{vaswani2017attention} so that each head can generate a different attention distribution and play a different role;
	\item applying label smoothing \cite{szegedy2016rethinking} to prevent the model from making over-confident predictions;
	\item integrating external LM which was trained with more text data \cite{kannan2018analysis};
	\item using minimum word error rate sequence discriminative training \cite{prabhavalkar2018minimum}.
\end{itemize}
With all these improvements, the final attention-based end-to-end system clearly outperformed the traditional hybrid system \cite{chiu2018state}. Different from CTC, a challenge to attention-based model is the attention was performed on top of the whole input utterance which means it cannot be performed in a streaming fashion even if the encoder can be done in a streaming mode.  
  

The frame-independence assumption in CTC  is the most criticized assumption as speech frames are correlated. On the other hand, the attention-based model has its drawback of not having monotonic left-to-right alignment and slow convergence. In \cite{kim2017joint}, the attention training is combined with CTC training in a multi-task learning way by using CTC objective function as the auxiliary function. Such a training strategy greatly improves the convergence of attention-based model and mitigates the alignment issue. In \cite{hori2017joint}, it was further advanced with jointly decoding with the scores from both attention-based model and CTC model. 


\section{Acoustic Model Robustness}\label{sec:robust}

Current state of the art systems can achieve remarkable recognition accuracy when the test and training sets match, esp. when both under quiet close-talk condition. However, the performance dramatically degrades under mismatched or complicated environments such as higher noise condition, including music or interfering talkers, or speech with strong accents \cite{Li14overview, Li15robust}. The solutions to this problem include adaptation, speech enhancement, and robust modeling.

\subsection{Acoustic Model Adaptation}\label{sec:adapt} 

In this section, we use speaker adaptation as an example scenario to describe acoustic model adaptation technologies. The same technology should be easily applied to the adaptation of new environments and tasks, etc. Typically the speaker independent (SI) models are trained from a large dataset with an objective to work best for all speakers. Speaker adaptation can significantly boost the performance of an individual speaker \cite{seide-2011, liao2013speaker}. However, we typically have limited adaptation data, and unsupervised adaptation is the main stream given prohibitive transcription cost. Current research focus is unsupervised adaption with limited amount of speaker-dependent data, which can be addressed with better adaptation criterion and model topology.  Since the adapted models are speaker dependent (SD), the size of the SD parameters is critical if we want to  scale to millions of speakers. This requires solutions to minimizing the SD model footprint while maintaining the adaptation  benefits. 

Given the limited amount of adaptation data, the SD model should not be far away from the SI model. \cite{yu2013kl} adds Kullback-Leibler divergence (KLD) regularization to the training criterion to prevent the adapted model from straying too far away from the SI model. This KLD  adaptation criterion has been proven very effective dealing with limited adaption data.  Most state-of-the-art SI models use senone (tied triphone states) as the output units. When limited amount of adaptation data is available, only very small amount of senones have been observed. In such a case, the adaptation turns to overfit the data distribution of these senones thus cannot generalize very well. In \cite{huang2015rapid}, a multi-task learning (MTL) framework was proposed by adding auxiliary monophone classification as the second task in addition to the primary seone classification task. As a result, the network adaptation is backed off to improving monophone classification accuracy when senones are not observed, hence increasing the generalization ability. 

In contrast to adjusting the adaptation criterion, most of works focus on how to use very small amount of parameters to represent speaker characteristics. One solution is the singular value decomposition (SVD) bottleneck adaptation \cite{Xue14} which produces low-footprint SD models by making use of the SVD-restructured topology \cite{Xue13}. The linear transformation is applied to each of the bottleneck layer by adding a $kXk$ SD matrices. The advantage of this approach is that only a couple of small matrices need to be updated for each speaker as $k$ is the low-rank value of the SVD reconstruction and usually is very small. This dramatically reduces the deployment cost for speaker personalization while producing more reliable estimate of the adapted model \cite{Xue14}. Works have been done to further reduce the size of the $kXk$ SD matrices. For example, when the adaptation data is very limited, the $kXk$ matrix can be reduced to a diagonal matrix, such as learning hidden unit contribution (LHUC) \cite{Swietojanski14adapt, swietojanski2016learning} and sigmoid adaptation \cite{zhao2015}. This is a tradeoff between the modeling capacity and generalization. The LHUC and sigmoid adaptation have much smaller number of adaptation parameters  compared to the SVD adaption, but they may not get similar accuracy improvement when the amount of adaptation data is increased. The observation that $kXk$ SD matrices usually are diagonally dominant matrices inspired the proposal of low-rank plus diagonal (LRPD) decomposition which decomposes the $kXk$ SD matrices into a diagonal matrix plus the multiplication of two low-rank matrices. By varying the low-rank values, the LPRD matrix generalizes the full-rank and the diagonal adaptation matrix, and hence can automatically utilize the adaptation data well instead of making tradeoff between model capacity and generalization. 

The subspace methods are another type of methods that also  aim to find a low dimensional subspace of the transformations, so that each transformation can be specified by a small number of parameters.
One popular method in this category is the use of auxiliary features, such as i-vector \cite{Saon13, senior2014improving}, speaker code \cite{Abdel13} , and noise estimate \cite{seltzer2013investigation} which are concatenated with the standard acoustic features. It can be shown that the augmentation of auxiliary features is equivalent to confining the adapted bias vectors into a speaker subspace \cite{yu2015adaptation}. Furthermore, networks can be used to  transform speaker features such as i-vectors into a bias to offset the speech feature into a speaker-normalized space \cite{miao2014towards}.  In addition to augmenting features in the input space, the  acoustic-factor features can also be appended in any layer of  deep networks \cite{Li14factorized}.

Other subspace methods include cluster adaptive training (CAT) \cite{tan2015cluster, wu2015multi} and factorized hidden layer (FHL) \cite{samarakoon2016factorized, Lahiru17}, where the transformations are confined into the speaker subspace.  Similar to the eigenvoice \cite{kuhn1998eigenvoices} or cluster adaptive training \cite{gales1998cluster} in the Gaussian mixture model era, CAT \cite{tan2015cluster, wu2015multi} in DNN training constructs multiple DNNs to form the bases of a canonical parametric space. During adaptation, an interpolation vector which is associated to a target speaker or environment is estimated online to combine the multiple DNN bases into a single adapted DNN. Because only the combination vector is estimated, the adaptation only needs very small amount of data for fast adaptation. However, this is again a tradeoff from the model capacity. In contrast with online estimation of the combination vector, \cite{delcroix2015context, delcroix2016context} directly uses the posterior vectors of the acoustic context to enable fast unsupervised adaptation. The acoustic context factor can be speaker, gender, or acoustic environments such as noise and reverberation. The posterior calculation can be either independent \cite{delcroix2015context} or dependent \cite{delcroix2016context} on the recognizer network. 

An issue in the CAT-style methods is that the bases are full-rank matrices, which require very large amount of training data. Therefore, the number of bases in CAT is usually constrained to few \cite{tan2015cluster, wu2015multi}. A solution is to use FHL \cite{samarakoon2016factorized, Lahiru17} which constrains the bases to be rank-1 matrices. In such a way, the training data for each basis is significantly reduced, enabling the use of larger number of bases. Also, FHL initializes the combination vector from i-vector for speaker adaptation, which helps to give the adaptation a very good starting point. In \cite{zhao2017eLRPD}, LRPD was extended into the subspace-based approach to further reduce the speaker-specific footprint in a very similar way to FHL.

\subsection{Speech Enhancement and Separation}\label{sec:enhance}

It is well known that the current ASR systems perform poorly when the speech is corrupted with heavy noise or interfering speech \cite{MonauralSpeechSepChallenge-Cooke2010,SingleChannelSep-Weng2015}. Although human listeners also suffer from poor audio signals, the performance degradation is significantly smaller than that in ASR systems. 

In recent years, many works have been done to enhance speech under these conditions. Although majority of the works are focused on single-channel speech enhancement and separation, the same techniques can be easily extended to multi-channel signals. 

In the monaural speech enhancement and separation tasks, it is assumed that a linearly mixed single-microphone signal ${y}[n]=\sum_{s=1}^{S} {x}_s[n]$ is known and the goal is to recover the $S$ streams of audio sources ${x}_s[n], s=1,\cdots,S$. If there are only two audio sources, one for speech and one for noise (or music, etc.) and the goal is to recover the speech source, it's often called speech enhancement. If there are multiple speech sources, it is often referred to as speech separation. The enhancement and separation is usually carried out in the time-frequency domain, in which the task can be cast as recovering the short-time Fourier transformation (STFT) of the source signals ${X}_s(t,f)$ for each time frame $t$ and frequency bin $f$, given the STFT of the mixed speech ${Y}(t,f)=\sum_{s=1}^{S}{X}_s(t,f)$. 

Obviously, given only the mixed spectrum ${Y}(t,f)$, the problem of recovering ${X}_s(t,f)$ is under-determined (or ill-posed), as there are an infinite number of possible ${X}_s(t,f)$ combinations that lead to the same ${Y}(t,f)$. To overcome this problem, the system has to learn a model based on some training set $\mathbb{S}$ that contains parallel sets of mixtures ${Y}(t,f)$ and their constituent target sources ${X}_s(t,f)$, $s=1, ... ,S$ \cite{SpeechSepTrainingTargets-wang2014,SpeechEnhanceWithDNN-xu2014,SpeechSepWithLSTM-weninger2015,JointMaskDNN-Huang2015,DeepClustering-hershey2015,DeepClustering2-isik2016,PIT-yu2017,PIT-Kolbak2017}. 

Over the decades, many attempts have been made to attack this problem. Before the deep learning era, the most popular techniques include computational auditory scene analysis (CASA) \cite{CASA-cooke2005,CASA-ellis1996,PerceptualCuesInCASA-wertheimer1938}, non-negative matrix factorization (NMF) \cite{sparseNMF-schmidt2006,NMF-SpeechSep-smaragdis2007,SparseNMF-le2015}, and model based approach \cite{IBM-SuperHuman-kristjansson2006,SpeechSepWithFactorialHMM-virtanen2006,SpeechSepWithAdaptedModel-weiss2007}, such as factorial GMM-HMM \cite{FactorialHMM-ghahramani1997}. Unfortunately these techniques only led to very limited success.

Recently, researchers have developed many deep learning techniques for speech enhancement and separation. The core of these techniques is to cast the enhancement or separation problem into a supervised learning problem. More specifically, the deep learning models are optimized to predict the source belonging to the target class, usually for each time-frequency bin, given the pairs of (usually artificially) mixed speech and source streams. Compared to the original setup of unsupervised learning, this is a significant step forward and leads to great progress in speech enhancement. This simple strategy, however, is still not satisfactory, as it only works for separating audios with very different characteristics, such as separating speech from (often challenging) background noise (or music) or speech of a specific speaker from other speakers \cite{JointMaskDNN-Huang2015}. It does not work well for speaker-independent multi-talker speech separation.

The difficulty in speaker-independent multi-talker speech separation comes from the label ambiguity or permutation problem. Because audio sources are symmetric given the mixture (i.e., $x_1+x_2$ equals to $x_2+x_1$ and both $x_1$ and $x_2$ have the same characteristics), there is no pre-determined way to assign the correct source target to the corresponding output layer during supervised training. As a result, the model cannot be well trained to separate speech.

Fortunately, several techniques have been proposed to address the label ambiguity problem \cite{SingleChannelSep-Weng2015,DeepClustering-hershey2015,DeepClustering2-isik2016,AtrractorNet4SpeechSeparation-chen2017,PIT-yu2017,PIT-Kolbak2017}. In Weng et al. \cite{SingleChannelSep-Weng2015} the instantaneous energy was used to solve the label ambiguity problem and a two-speaker joint-decoder with a speaker switching penalty was used to separate and trace speakers. This work achieved the best result on the dataset used in 2006 monaural speech separation and recognition challenge \cite{MonauralSpeechSepChallenge-Cooke2010}. However, energy, which is manually picked, may not be the best information to assign labels in all conditions. Actually, we have found that in many cases pitch difference is a more important cue. In Hershey et al. \cite{DeepClustering-hershey2015,DeepClustering2-isik2016} a novel technique called deep clustering (DPCL) was proposed. In this model, it is assumed that each time-frequency bin belongs to only one speaker. During training, each time-frequency bin is mapped into an embedding space. The embedding is so optimized that time-frequency bins belong to the same speaker are closer and that of different speakers are farther away in this space.  During evaluation, a clustering algorithm is used upon embeddings to generate a partition of the time-frequency bins. To further improve the performance, they stacked yet another network to estimate real masks for each source stream given the results from the deep clustering \cite{DeepClustering2-isik2016}. Chen et al. \cite{AtrractorNet4SpeechSeparation-chen2017} proposed a technique called deep attractor network (DANet). Following DPCL, their approach also learns a high-dimensional embedding of the acoustic signals. Different from DPCL, however, it creates cluster centers, called attractor points, in the embedding space to pull together the time-frequency bins corresponding to the same source. The main limitation of DANet is the requirement to estimate attractor points during evaluation time.

In Yu et al. \cite{PIT-yu2017} and Kolbak et al.\cite{PIT-Kolbak2017}, a simpler technique named permutation invariant training (PIT) was proposed to attack the speaker independent multi-talker speech separation problem. In this new approach, the source targets are treated as a set (i.e., order is irrelevant). During training, PIT first determines the output-target assignment with the minimum error at the utterance level based on the forward-pass result. It then minimizes the error given the assignment. This strategy elegantly solves the label permutation problem and speaker tracing problem in one shot. Unlike other techniques such as DPCL and DANet that require a separate clustering step to trace speech streams during evaluation, PIT does not require a separate tracing step (and thus can be used in real-time systems). Instead, each output layer is corresponding to one stream of sources. In PIT the computational cost associated with label assignment is negligible compared to the network forward computation during training, and no label assignment (and thus no cost) is needed during evaluation. Recently, Hershey et al. \footnote{based on personal communication} have found out that in DPCL the embeddings actually are grouped into two classes, instead of many different classes for different speakers, in the two-speaker separation problem. This indicates that DPCL essentially learns separation models that is very similar to that learned by PIT. DPCL, DANet, and PIT all achieve similar performance on speaker-independent two- to three-talker speech separation tasks yet PIT is the simplest among all, can be used in real-time systems, and can be easily combined with other techniques. Moreover, unlike DPCL or DaNet, PIT does not need to know or estimate the number of streams in the mixture. We therefore believe that PIT is most promising among these techniques. Similar to progresses made when converting the speech separation problem from an unsupervised learning problem into a supervised learning problem, PIT converts the speech separation problem from supervised learning with ambiguous label to that with clear labels.

For speech recognition, we can feed each separated speech stream to ASR systems. Even better, the deep learning based AM may be jointly optimized end-to-end with the separation component, which is often an RNN. Since separation is just an intermediate step, Yu et al. \cite{PITASR-yu2017} proposed to directly optimize the cross-entropy criterion against senone labels using PIT without having an explicit speech separation step. Their preliminary results on AMI dataset indicate that PIT can significantly improve the recognition accuracy, compared to the models trained with single-talker speech,  when recognizing two-talker mixed speech. In \cite{chen2018progressive}, PIT-ASR was further advanced by imposing a modular structure on the network, applying progressive pretraining, and improving the objective function with teacher-student learning and a discriminative training criterion.
Even using the ASR criterion to guide the separation of multi-speaker as PIT-ASR, current multi-speaker ASR system still relies on signals from source speakers to get the time alignment. Hence the multi-speaker mixture training data indeed is artificially generated. In \cite{E2EmultiASR2018}, an attempt was done by using end-to-end ASR training criterion for multi-speaker ASR. In this way, only a pretrained separation network is required, and then the whole system can be optimized with only transcription-level labels without using the source signal from each speaker. This opens a door for using real multi-speaker mixture data for training. 

The speech separation methods discussed so far rely on spectral information for the separation. Significant performance improvement has been obtained with the recently proposed multi-channel separation methods which utilize spatial information in addition to spectral information \cite{drude2017tight, chen2017cracking, chen2018EFFICIENT}. In \cite{drude2017tight}, clustering was performed using spatial feature in addition to the embedding features derived from deep clustering, which requires a complex clustering algorithm to deal with these two types of features.  In \cite{chen2017cracking}, a set of fixed beamformers is first applied to an observed multi-channel signal to yield a set of beamformed audio. Then a separation network is applied to each beamformed signal. Out of the speech separation results for all the beams, the best separated signal is selected for each speaker by a post selection algorithm. This system  significantly improved the state of the art in speech separation, obtaining comparable performance to a minimum variance distortionless response beamformer that uses oracle location, source, and noise information. However, performing speech separation on each of the beamformed signals is computationally very expensive. The system computational cost was then significantly reduced by first using a beam selection network to predict the best beam for each speaker, and then feeding the outputs from the selected beams into a PIT network for further separation \cite{chen2018EFFICIENT}. 

\subsection{Robust Training}\label{sec:robusttrain} 

The success of deep neural networks relies on the availability of a large amount of transcribed data to train millions of model parameters. However, deep models still suffer reduced performance when exposed to test data from a new domain. Because it is typically very time-consuming or expensive to transcribe large amounts of data for a new domain, domain-adaptation approaches have been proposed to bootstrap the training of a new system from an existing well-trained model \cite{liao2013speaker,Xue14}. These methods still require transcribed data from the new domain and thus their effectiveness is limited by the amount of transcribed data available in the new domain. Although unsupervised adaptation methods can be used by generating labels from a decoder, the performance gap between supervised and unsupervised adaptation is large \cite{liao2013speaker}. 

Recently, the concept of adversarial training \cite{goodfellow2014generative} was explored for noise-robust ASR  \cite{shinohara2016adversarial, serdyuk2016invariant, Sun17GRL, dsn_meng, Meng18GRLTS, Yi18GRL, Sun18GRL, Liu18GRL} and speaker invariant training \cite{saon2017english, Meng18SIT}. This solution is a pure unsupervised domain adaptation method without utilizing too much knowledge about the new domain. The domain can be clean or noisy environment for noise-robust ASR, and different speakers for speaker invariant training. The idea is to have three networks in the model: the encoder network, the recognizer network, and the domain discriminator network. The encoder network generates the intermediate representation, which will be used in both the recognizer network to generate posteriors of phoneme units and the domain discriminator network to generate domain labels. The intermediate representation is learned adversarially to the domain discriminator, i.e., to minimize the domain classification accuracy. In such a way, the intermediate representation is invariant to the input of different domains. At the same time, the intermediate representation is trained to maximize the phoneme classification accuracy with the source domain labels. During testing time, the encoder network generates the intermediate representation from the target domain data and input it into the recognizer network. The training is done by inserting a gradient reverse layer (GRL) \cite{ganin2014unsupervised} between the encoder network and the domain discriminator network. During forward propagation, GRL acts an identity transform. During back propagation, GRL takes the gradient from the domain discriminator network, multiples it by a negative constant, and then passes it to the encoder network. 

The advantage of GRL-based unsupervised adaptation is that it doesn't require any knowledge about the target domain. In contrast, the adaptation to the new domain should be more effective if we have some domain knowledge by simulating the target domain data. For example, if the source domain is a clean environment and the target domain is a noisy environment, we can simulate noisy environment data and then do the multi-style training \cite{lippmann1987multi} with simulated data, e.g., in \cite{kostudy}. However, the multi-style training does not use the well-trained source model which usually has very high accuracy in the source domain. 

Recently, the teacher/student (T/S) learning \cite{Li17TS} method was proposed to perform adaptation without the use of transcriptions.  The data from the source domain are processed by the source-domain model (teacher) to generate the corresponding posterior probabilities or soft labels. These posterior probabilities are used in lieu of the usual hard labels derived from the transcriptions to train the target (student) model with the parallel data from the target domain. With this approach, the network can be trained on a potentially enormous amount of training data and the challenge of adapting a large-scale system shifts from transcribing thousands of hours of audio to the potentially much simpler and lower-cost task of designing a scheme to generate the appropriate parallel data.  

Even with the same amount of data, T/S learning with soft labels sometimes was shown to outperform CE training with hard labels. This is because soft labels provide much more information than hard labels, making the training easier in challenging tasks, such as ASR in noisy environments. Soft labels indicate how the teacher model view the world, i.e., data samples. For example, if the target phone (hard label) is “ah”, the teacher model usually has a very high probability with “ah”. At the same time, it may also have reasonable probabilities for vowels such as “ax” and “ae” etc., but has very low probabilities for consonants such as “v”, “b” etc.  This information is very helpful when training models in challenging environments because the network cannot predict the target “ah” perfectly and predicting other phonemes as 0 as what the hard label requires (i.e., there is no differentiation between vowels except target “ah” and consonants – this is hard to learn). In contrast, the T/S learning gives  a very relaxed and easy task: predict vowels with relative higher probability and consonants with low probability.

The T/S learning approach is closely related to other approaches for adaptation or retraining that employ knowledge distillation \cite{hinton2015distilling}. In these approaches, the soft labels generated by a teacher model are used as a regularization term to train a student model with conventional hard labels. For example, knowledge distillation was used to train a system on the Aurora 2 noisy digit recognition task, using the clean and noisy training sets \cite{markov2016robust}. In \cite{Watanabe17} it was shown that for the multi-channel CHiME-4 task, soft labels could be derived using enhanced features generated by a beamformer then processed through a network trained with conventional multi-style training. In all cases, the soft labels provided by the teacher network regularized the conventional training of the student network using hard labels derived from transcriptions. Thus, the use of additional unlabeled training data was not possible. 

In conclusion, domain adaption without labeled data will be an important research direction. If we don't have any knowledge about the target domain, adversarial training should be a good way to go. On the other hand, if we can simulate data similar to the target domain data, T/S learning and knowledge distillation are good methods. Especially, the T/S learning methods forgoes the need for hard labels from the data in the new domain entirely and relies solely on the soft labels provided by the parallel corpus and well-trained source model. This allows the use of a significantly larger set of adaptation data which adds robustness to the resulting model. In \cite{Li2018Speaker}, T/S learning was successfully used  to develop a far-field speaker system by leveraging tens of thousands of hours unlabeled data.   

\section{Acoustic Models with Efficient Decoding}\label{sec:efficient}  

Training deep networks by stacking multiple layers helps to improve WER. However, the computational cost becomes a concern, especially to the industry deployment where realtime is always with high priority. There are several ways to reducing the runtime cost. 

The first one is to use singular value decomposition (SVD) which was originally proposed in \cite{Xue13} and has been widely used. The SVD method decomposes a full-rank matrix into two lower-rank matrices, hence can  significantly reduce the number of parameters in deep models without losing accuracy after retraining.  This is general to any deep network structure. In \cite{lu2016learning, prabhavalkar2016compression}, a similar method was proposed for learning compact LSTMs via low-rank factorization and parameters sharing schemes.  

The second way is to employ teacher-student (T/S) learning or knowledge distillation, such as the works in \cite{chan2015transferring, geras2015compressing, lu2017knowledge, Cui2017knowledge}. T/S learning was proposed in \cite{li2014learning} to compress a standard DNN model by minimizing the KLD between the output distributions of the small-size DNN and a standard large-size DNN. The learning equals to the CE training using the soft label generated by the teacher model as the target for the student learning. The concept of T/S learning was extended as the concept of distilling the knowledge in \cite{hinton2015distilling} by combing the CE training using soft labels with the standard CE training using the 1-hot vector as the target. The soft target in knowledge distillation serves as the regularization term to the standard CE training.  Recently, Microsoft and IBM continuously broke the WER record on the Switchboard task \cite{xiong2016achieving, saon2017english}. The built systems are usually giant models ensemble of multiple deep models. Such system cannot be deployed to realtime application. In such a scenario, T/S learning or knowledge distillation provides a good solution to getting a compact model with high modeling capacity. A most recent success of T/S compression is in developing a far-field device key word spotting (KWS) CTC model \cite{Li2018Speaker} with almost the same KWS but with only 1/27 footprint of a large-size KWS CTC model. Note that T/S learning is essentially a frame-by-frame CE training criterion with soft labels, it may be desirable to develop a sequence-based T/S criterion \cite{wong2016sequence, kanda2017investigation} if the teacher model was trained with sequence criterion although in practice the standard T/S learning works very well even in such a situation \cite{Li2018Speaker}. 

The third method is to compress the models by heavy quantization,  applying either very low-bit quantization or vector quantization. \cite{vanhoucke2011improving} gives very nice summary of technologies to speed up the runtime evaluation of deep networks. Those technologies including 8-bit quantization don't require re-training deep networks. However, the ASR accuracy is significantly lost when the model is compressed heavily into even lower bits or the network structure becomes more complex. Therefore, refining with quantization is important for both very low-bit quantization \cite{alvarez2016efficient, takeda2017acoustic} and vector quantization \cite{wang2015small} so that the training and testing are consistent.  

The fourth solution is to work on model structures. LSTM with projection layer (LSTMP) was proposed to reduce the computational cost by adding a linear projection layer after the LSTM layer\cite{Sak2014long}. This projected vector has lower dimension than the output of the LSTM layer, and is used to replace the recurrent input $\textbf{h}_{t-1}$ to the LSTM unit. Although the initial purpose of LSTMP is to reduce runtime cost, it was also reported helpful to error rate reduction \cite{Sak2014long, Miao15} which is possibly because parameter reduction helps to generalize better if the LSTM model is too strong. The projection layer is not used in the later Google's CTC work \cite{sak2015learning, sak2015fast, Senior15} where the number of memory cells in LSTM is smaller and the amount of training data is 20 times larger.  In \cite{Miao16}, the runtime cost of LSTM is reduced with model simplification by coupling input gates and forget gates.

Finally, the correlation across frames can be used to reduce the evaluation frequency of deep network scores. For DNNs or CNNs, this can be done with a frame-skipping strategy by computing the acoustic scores once every several frames and copying the acoustic scores to the frames in which no acoustic score is evaluated during decoding \cite{vanhoucke2013multiframe}. However, for decoding with LSTM models,  the same frame-skipping strategy needs to be done in both training and testing stages so that the behaviors of LSTM's memory units are consistent \cite{Miao16}. Recently, LFMMI \cite{povey2016} and lower frame rate (LFR) LSTM-RNN \cite{pundak2016lower} were proposed  to work on the 30ms units instead of the traditional 10ms units by modeling phonemes instead of states. Because of using larger input units, the LFMMI and LFR models only need 1/3 computation cost of the traditional networks which operate on 10ms inputs. As CTC uses phones or even words as the output targets, it can also work on 30ms or even large inputs, therefore significantly reduce the runtime cost. 

\section{Future Directions}\label{sec:future}

The research frontier has been shifted from close-talk microphone to far-field microphone, driven by increased demand from users to interact with devices without wearing or carrying a close-talk microphone. For example, Amazon's Echo and Google Home have now been deployed in many families across the world. Many difficulties hidden by close-talk microphones now surface when far-field microphones are used. This is because in the far-field scenario, the energy of speech signal is very low when it reaches the microphones. Comparatively, the interfering signals, such as background noise, reverberation, and speech from other talkers, become so distinct that they can no longer be ignored.

Although many of the speech recognition techniques developed for close-talk microphones can be directly applied to far-field microphones, these techniques show inferior performance under distant recognition scenario. To ultimately solve the distant speech recognition problem, we need to optimize the whole pipeline starting from audio capturing (e.g., microphone-array signal processing) to acoustic modeling and decoding. We perceive following possible research directions.

First, although we have made some interesting progresses in monaural speech enhancement, separation, and recognition, much more improvements are desired. For example, the current PIT system performs very well when separating speech mixtures of different genders. However, the separation quality is deteriorated when two same-gender speakers speak at the same time. Will the quality be good enough if we also exploit beam-forming results and multi-channel information? Is there a better model than conventional LSTMs for speech separation and tracing? How can we exploit additional information, such as language model, by feeding information from the decoder back to the speech enhancement and separation component and by jointly considering all streams of speech when making decoding decision?

Second, the end-to-end optimization strategy is desired, given its simplicity and joint optimization characteristics, if we only need to optimize for the decoding result and have sufficient training data. This has been proven effective with word-based CTC when trained with hundreds of thousands hours data.  However, it is not feasible to get that large amount of data for most tasks. Considering that current end to end systems trained with thousands hours of data use two LMs, one that implicitly built-in and trained with the AM, and one that is separated and trained using text data, we would guess that further accuracy improvement can be achieved if we can integrate the second LM into the end to end system and optimize them jointly when audio signals are available and separately when only text data is available. If this turns out to be beneficial, would it affect our choice of modeling unit in end to end systems? For example, if Chinese characters are used as modeling unit, adding a new character would be difficult, partly because it will change the number of classes in the model and partly because there may not be enough data to train the new character. However, since the pronunciation of the new character would be one of the syllables that have already been covered by other characters, adding new characters can be extremely easy if syllable is the modeling unit since the pronunciation knowledge can be directly transfered. To further extend it, can we design the model so that components are built and can be transferred at different scales? Both the CTC and attention-based model have their individual pros and cons. The joint training with two models together is first but superficial step toward better end to end modeling. Can we formulate a single model by taking the advantages of both models? 

Third, even the models trained with huge amount of data are lack of robustness. This is because training-test mismatch is unavoidable given the cost of data collection. Adding simulated data can alleviate the problem if we can foresee the possible variations but may still be not sufficient. Can we design a model (e.g., as a special nonlinear dynamic system) that constantly adapts itself within some limit, e.g., controlled by some kernel size? Can such models automatically exploit information gained from past similar speakers to quickly adapt to new speakers? It is even more interesting if such models can gradually identify the most reliable regularities in the data.


\ifCLASSOPTIONcaptionsoff
  \newpage
\fi



%

\bibliographystyle{IEEEtran}
\bibliography{mybib}


%


\end{document}